\documentstyle[epsf,graphics,epsfig]{mn}

\newcommand{\Msolar}{\mbox{\,$\rm M_{\odot}$}}        
\newcommand{\Mblack}{\mbox{\,$\rm M_{bh}$}}     
\newcommand{\Mbulge}{\mbox{\,$\rm M_{bulge}$}}

\newcommand{\spsc}[2]{\mbox{$\rm #1^{#2}$}}     
\newcommand{\sbsc}[2]{\mbox{$\rm #1_{#2}$}}     
\newcommand{\ang}{\mbox{$\rm \AA$}}

\title[AGN blackhole masses]{The black-hole masses of Seyfert galaxies and 
quasars}
\author[R.J. McLure \& J.S. Dunlop]
{R.J. McLure$^1$ \& J.S. Dunlop$^2$\\
 $^{1}$Nuclear and Astrophysics Laboratory, University of Oxford,
Keble Road,Oxford, OX1 3RH\\
$^{2}$Institute for Astronomy, University of Edinburgh. Royal
Observatory, Edinburgh EH9 3HJ.\\}
\date{Submitted for publication in MNRAS}
\begin{document}
\maketitle
 
\begin{abstract}
The central black-hole masses of a sample of 30 luminous quasars are
estimated using $\sbsc{H}{\beta}$ {\sc fwhm} measurements from a 
combination of new and previously-published nuclear
spectra. The quasar black-hole mass estimates are combined with
reverberation-mapping measurements for a sample of Seyfert galaxies
(Wandel, Peterson \&\ Malkan 1999) in order to study AGN
black-hole masses over a wide range in nuclear
luminosity. The link between bulge luminosity and black-hole mass is
investigated using two-dimensional disc/bulge decompositions of
the host-galaxy images, the vast majority of which are high
resolution {\sc hst} observations.  It is found that black-hole mass
and bulge luminosity are well correlated and follow a relation
consistent with that expected if black-hole and bulge mass are
directly proportional. Contrary to the results of Wandel (1999) 
no evidence is found
that Seyfert galaxies follow a different $\Mblack-\Mbulge$ 
relation to quasars. However, the black-hole mass distributions of
the radio-loud and radio-quiet quasar sub-samples are found to be
significantly different, with the median black-hole mass of the
radio-loud quasars a factor of three larger than their radio-quiet
counterparts. Finally, utilizing the elliptical galaxy 
fundamental plane to provide
stellar velocity dispersion estimates, a comparison is performed
between the virial $\sbsc{H}{\beta}$ black-hole mass estimates and those of
the $\Mblack-\sigma$ correlations of Gebhardt et al. (2000a) and
Merritt \&\ Ferrarese (2000). With the disc geometry of the 
broad-line region adopted in this paper,
the virial $\sbsc{H}{\beta}$ black-hole masses indicate that the correct
normalization of the black-hole vs. bulge mass relation 
is $\Mblack\simeq0.0025\Mbulge$, while the standard assumption of
purely random broad-line velocities leads to
$\Mblack\simeq0.0008\Mbulge$. The normalization of
$\Mblack\simeq0.0025\Mbulge$ provided by the disc model is in
remarkably good agreement with that inferred for our quasar sample
using the (completely independent) $\Mblack-\sigma$ correlations.

\end{abstract}
\begin{keywords}
galaxies: active -- galaxies: nuclei -- galaxies: Seyfert -- quasars: general 
\end{keywords}
 
\section{Introduction}
The evidence that supermassive black-holes are ubiquitous in nearby
inactive galaxies is now extremely strong (eg. van der Marel 1999,
Kormendy \&\ Richstone 1995). Furthermore, several
studies have concluded that the masses of these dormant black-holes
are directly proportional to the mass of the galaxy
bulge (spheroidal) component (eg. Kormendy \&\ Richstone 1995, Magorrian et
al. 1998). Although a significant scatter is present in the
correlation ($\Delta \Mblack \simeq 0.5$ dex at a given bulge mass), the
data appear to suggest that
$\Mblack=(0.002\rightarrow0.006)\Mbulge$. 
While studies of nearby inactive galaxies have seen a reasonably
coherent picture emerge, the situation with regard to active galaxies is
less certain. Although the black-hole mass relation observed in nearby
inactive galaxies can be comfortably reproduced by 
combined galaxy+quasar formation models (eg. Wilman, 
Fabian \&\ Nulsen 2000, Kauffmann \&\ Haehnelt 2000), observational
studies to investigate the form of the $\Mblack-\Mbulge$ relation in 
active galaxies have led to seemingly contradictory results. 

In a study of the $\Mblack-\Mbulge$ relation in quasars, Laor (1998)
 used the {\sc fwhm} of the broad $\sbsc{H}{\beta}$ emission line, together
with a calibration between the broad-line region (BLR)
 radius and quasar UV luminosity, to obtain
virial black-hole mass estimates for 14 PG quasars. Using estimates of
 the host-galaxy bulge luminosities from the {\sc hst} study of Bahcall et
al. (1997), Laor found a $\Mblack-\sbsc{L}{host}$ correlation which 
agreed reasonably well with the results of Magorrian et al. (1998). 
In contrast, Wandel (1999) found that virial black-hole mass estimates
 for a sample of 17 Seyfert1 galaxies, based on the $\sbsc{H}{\beta}$
 {\sc fwhm} and reverberation mapping measurements of the BLR radius, were
substantially smaller than would be expected from the
 $\Mblack-\Mbulge$ relation seen in nearby inactive galaxies.
 Using estimates of the Seyfert galaxy bulge luminosities
from the empirical morphology-based formula of Simien \&\ de
Vaucouleurs (1986), Wandel found that $\Mblack/\Mbulge\sim10^{-3.5}$, 
a factor of between ten and twenty lower than found for both nearby 
inactive galaxies (Magorrian et al. 1998) and for luminous 
quasars (Laor 1998).

The implication of the Wandel (1999) result is that there exists a
difference in the formation or triggering mechanism at work in
Seyfert galaxies and quasars. This possibility is explored by Wang,
Biermann \&\ Wandel (2000) who present a model in which 
$\Mblack/\Mbulge \propto \sigma^{1.4}$, where $\sigma$ is the velocity
dispersion of the accreting gas. In this model the high values of
$\Mblack/\Mbulge$ found in quasars and nearby bulges
are the asymptotic, high-velocity case, possibly produced by a violent
merger event. In contrast, the much lower ratios found in Seyfert galaxies
are the result of the accretion of lower velocity gas triggered by
tidal disruption.

In this paper the $\Mblack-\Mbulge$ relation is examined over a wide
range in AGN luminosity by combining new results for a sample of 
30 luminous (M$_{\rm V}<-23$) quasars with a re-analysis of the 
bulge masses attributed to the Wandel (1999) Seyfert galaxy sample.
First, in Section 2 we briefly review the complications and potential
uncertainties in deriving black-hole mass estimates from emission-line
widths, and explore how best to model the geometry of the broad-line region 
for the purpose of black-hole estimation in AGN.
Next, in Section 3 we describe the sample of objects selected for study,
present the new spectroscopic observations, and explain how reliable
bulge luminosities have been extracted from existing {\sc hst} images.
In Section \ref{main} three important questions are addressed. 
Firstly, do powerful AGN actually display a $\Mblack-\Mbulge$ 
correlation consistent with that found in nearby 
inactive galaxies? Secondly, do the apparent differences between
Seyfert galaxies and quasars actually stem from fundamental
differences between the two AGN classes, or simply from systematic errors
in estimating Seyfert bulge luminosities? And finally, do differences in
black-hole mass and gas accretion rates play a crucial role in
determining the radio properties of AGN? In Section \ref{fp} stellar 
velocity estimates for a sub-sample of the quasars are used to compare 
our virial $\sbsc{H}{\beta}$ black-hole mass estimates with those of 
the $\Mblack-\sigma$ relations discovered by Gebhardt et al. (2000a) 
and Merritt \&\ Ferrarese (2000). Finally, our main conclusions are
summarized in Section 6. Unless otherwise specified
all cosmological calculations performed in this paper assume 
$\sbsc{H}{0}=50$ kms$^{-1}$ Mpc$^{-1}$, q$_{0}=0.5$ and $\Lambda=0$.

\section{The virial black-hole mass estimate}
\begin{figure*}
\centerline{\epsfig{file=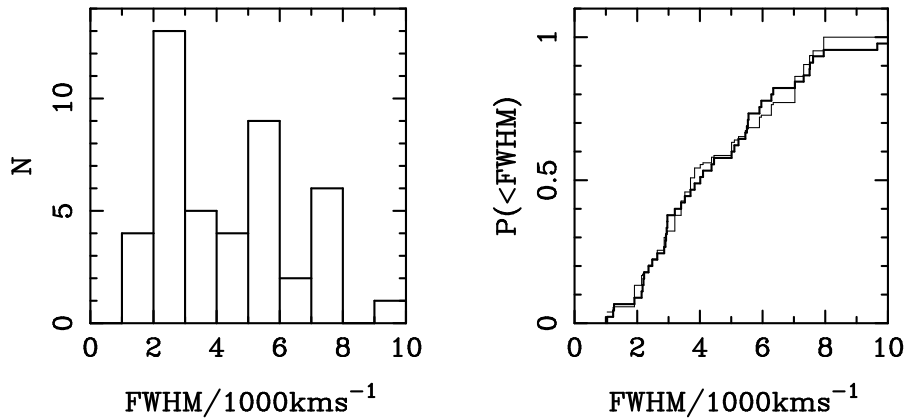,width=12.0cm,angle=0,clip=}}
\caption{The left-hand panel shows the distribution of
$\sbsc{H}{\beta}$ {\sc fwhm} measurements for the 45 objects in the
combined quasar+seyfert sample. The right-hand panel shows the 
cumulative {\sc fwhm}
distributions displayed by the data (thick line) and that of the disc
BLR model discussed in the text (thin line). The two cumulative 
distributions can be seen to be extremely similar, with the KS test 
probability of $p=0.99$ showing them to be statistically indistinguishable.}  
\label{fwhmfig}
\end{figure*}
The basic theory of obtaining virial black-hole mass estimates from quasar
emission lines has been discussed extensively in the literature
(eg. Wandel, Peterson \&\ Malkan 1999) and consequently
only a brief outline is provided here. The underlying assumption is
that the broad emission lines are produced by material which is
gravitationally bound and orbiting with keplerian velocities. If this
assumption is applicable then an estimate of the central mass is
given by $ \Mblack= \sbsc{R}{BLR} \spsc{V}{2} \spsc{G}{-1} $, 
where $\sbsc{R}{BLR}$ is the radius of the BLR and V is the velocity 
of the line-emitting material. There are two standard methods of
estimating the radius of the broad-line region. 
The first of these, and presumably the
more accurate, is to use the time delay between continuum and
line variations. For the sample of 
Seyfert galaxies studied by Wandel, Peterson \&\ Malkan (1999) and 
Wandel (1999), long-term monitoring has provided such reverberation 
mapping estimates of the radius of the broad-line region. In the absence of 
reverberation mapping data for the quasar sample it is necessary to 
use a more indirect estimate of the broad-line radius. 
The method used here is the correlation between $\sbsc{R}{BLR}$ 
and monochromatic luminosity at $5100\ang$ found by Kaspi et
al. (2000), from their reverberation 
mapping results for 17 PG quasars. When converted to the cosmology 
adopted in this paper, the Kaspi et al. correlation becomes: 
\begin{equation}
\sbsc{R}{BLR}=18.65 \left( \frac{\lambda L_{\lambda}(5100\ang)}{10^{37}W}^{0.7}
\right) {\rm lt-days}
\end{equation}

The other necessary component for the virial mass estimate is a measure
of velocity of the line emitting material. To obtain this, Wandel (1999)
adopted the assumption of random orbits, in which case V
$=\frac{\sqrt{3}}{2}$ {\sc fwhm}. However, the assumption of random orbits seems
unrealistic for quasars. In particular, for radio-loud quasars where the
radio core:lobe ratio provides an independent measure of orientation,
there exists strong evidence that the velocity-field of the broad-line region 
is better represented by a combination of a random isotropic component,
with characteristic velocity $\sbsc{V}{r}$, and a component only in the plane of 
the disk, with characteristic velocity $\sbsc{V}{p}$ (Wills \& Browne 1986). 

In this case, the observed {\sc fwhm} will be given by 
\begin{equation}
{\rm \sc{fwhm}} = 2(\sbsc{V}{r}^2 + \sbsc{V}{p}^2{\rm sin}^2\theta)^{1/2}
\end{equation}
\noindent
where $\theta$ is the angle between the disc normal, and the line
of sight to the observer. In the case of radio-quiet objects we have 
no guide as to what precise
value of $\theta$ should be adopted for a particular object. However, 
attempts to unify radio-loud quasars and galaxies (e.g. Barthel
1989), and imaging polarimetry studies of radio-quiet AGN (e.g. Antonucci
\&\ Miller 1985) support a picture in which an active nucleus is likely to
appear to the observer as a quasar provided $\theta < 45^{\circ}$. 

In
order to test the viability  of this disc-model we have attempted to
reproduce the form of the cumulative {\sc fwhm} distribution of the combined
quasar+seyfert sample. As can be seen from Fig 1b., the cumulative
distribution of the data is convex in shape. This form of distribution
cannot in fact be produced by a single disc model in which the angle
between the disc normal and observer's line of sight is assumed to be
randomly distributed between $\theta=0$ and some adopted upper limit
$\theta=\sbsc{\theta}{max}$. The reason for this is simply that, 
given a random distribution of theta (and
$\sbsc{V}{p}$ significantly larger than $\sbsc{V}{r}$) the cumulative distribution
of {\sc fwhm} grows proportional to solid angle $1-\cos{\theta}$.

However, we have found that the form of the observed distribution
can in fact be almost perfectly reproduced by a model which includes only
one additional free parameter. Specifically we have relaxed the
assumption of
a purely random distribution of theta for $\theta < \sbsc{\theta}{max}$ by
exploring whether the data can be described by a combination of two
randomly-oriented disc-model
populations with the same characteristic orbital velocity but
different angular constraints. The best-fitting model parameters 
were found to be $\sbsc{V}{p}=5500$ 
km$s^{-1}$, $\sbsc{\theta}{max}(1)=46^{\circ}$, 
$\sbsc{\theta}{max}(2)=20^{\circ}$, with the two models contributing
 $58\%$ and $42\%$ of the combined model distribution respectively. In
 both cases the model AGN are assumed to lie randomly at angles between 
$0<\theta<\sbsc{\theta}{max}$ to the line of sight. Although
 two-component disc-models with non-zero random velocity components
 and independent orbital velocity components were considered, it was
 found that the introduction of these extra free parameters did not
 produce significantly improved fits. The best-fitting model
 distribution is shown as the thin line in Fig 1b. and can be seen to 
be an excellent representation of the data, with an application of the
 Kolmogorov-Smirnov test showing the two distributions to be
 indistinguishable ($p=0.99$). By substituting the fitted model 
parameters into Eqn 2. it is now possible
to determine the average relationship between observed {\sc fwhm} and actual
orbital velocity $\sbsc{V}{p}$ by calculating $<\sin{\theta}>$
as weighted by solid angle. We therefore find that on average
$\sbsc{V}{p}=1.5\times$ {\sc fwhm}, and it is this relation which is adopted
throughout the remainder of this paper. 

As can be seen from the best-fitting parameters presented above, the
dominant population of objects in this two-component disc model of the
BLR is constrained to lie within $\sim45^{\circ}$ to the line of sight, as
expected in the unified model (eg. Barthel 1989). In contrast, the second
population of objects is required to lie within a smaller range of
angles to the line of sight
than is normally assumed in the unified model
(ie.$\sim 20^{\circ}$). However, one can argue that this is entirely 
reasonable given that the present sample is largely dominated by
powerful, optically-selected broad-line AGN. However, it is worth
noting the implications of the assumptions in this simple model,
and the effect of relaxing them. Due to the fact that the virial
black-hole mass estimate is proportional to $\sbsc{V}{p}^{2}$, the
adoption of the standard model in which BLR velocities are purely
random ($\sbsc{V}{p}=0.87\times${\sc fwhm}) leads to
black hole mass estimates three times {\it smaller} than derived in this
paper. Further evidence in support of the appropriateness of the BLR
model adopted here is discussed in Section 4.1 and Section 5.

\section{the sample, observations and data reduction}

The sample of 30 luminous quasars (M$_{\rm V}<-23$) studied in this
paper consists of two optically-matched sub-samples of 17 
radio-quiet quasars and 13 radio-loud quasars. 
The unique feature of this sample is that all of its
members have accurate bulge luminosities available from detailed
two-dimensional modelling of {\sc hst} images. Nineteen of the objects
are taken from the sample studied by McLure et al. (1999) and Dunlop
et al. (2000), with a further eleven taken from the sample of Bahcall
et al. (1997). This is therefore the best dataset currently available
for this type of investigation. The sample of Seyfert
galaxies consists of fifteen of the seventeen objects
studied by Wandel (1999). The two objects which are not included, Akn
120 \&\ Mrk 110, have been excluded on the basis that no suitable {\sc
hst} archive data exists for these objects, and that a reliable 
disc/bulge decomposition could not be identified in the literature. 

\subsection{Nuclear spectra}

The new observations presented in this paper consist of nuclear
spectra of 13 of the quasar sample which were obtained in December 1999 
on the 2.5m Isaac Newton Telescope on La Palma. All observations used the
Intermediate Dispersion Spectrograph (IDS) with the 235mm camera and the
EEV10 $4096\times 2048$ pixel CCD as the detector. 
Integration times on individual objects were $\ge3000$ seconds,
 with the RV300 grating providing spectra centred 
on $5800$\AA\, with a resolution of $\simeq4.5$\AA. At the average
redshift of the quasar sample ($z\sim0.2$) this ensured that the
spectra were always approximately centred on the H$_{\beta}$
line, with the wavelength coverage of $\sim4000$\AA\, allowing 
an accurate continuum determination. 

The reduction of the spectra was performed in {\sc iraf} in the standard
way, including flux calibration, and correction for atmospheric
extinction and galactic reddening. The final wavelength calibrated
spectra were interpolated onto a linear dispersion of 1.5\AA/pix and
are shown in Fig \ref{spectra}.

The H$_{\beta}$ {\sc fwhm} measurements for 16/17 of the remaining quasars in
the sample are taken from the Boroson \&\ Green (1992) and Corbin
(1997) studies. Given that the H$_{\beta}$ {\sc fwhm} enters into the virial
mass estimate quadratically, it is obviously important that there is
no large systematic offset between the two sets of spectra. 
Fortunately, five of the quasars for which new spectra have been
obtained also feature in the Boroson \&\ Green study.
 For these five objects the {\sc fwhm} values
determined from the new spectra agree well with those of Boroson \&\
Green, $\Delta{\sc fwhm}=10\pm4\%$, providing confidence that large
variations in measured {\sc fwhm} values do not introduce large
systematic errors into the virial black-hole estimates. The
$\sbsc{H}{\beta}$ {\sc fwhm} measurement for one final object,
1635+119, is taken from the spectrum of Wills \&\ Browne (1986).

\subsection{Host-galaxy bulge luminosities}
\label{bulge}
\begin{figure*}
\centerline{\epsfig{file=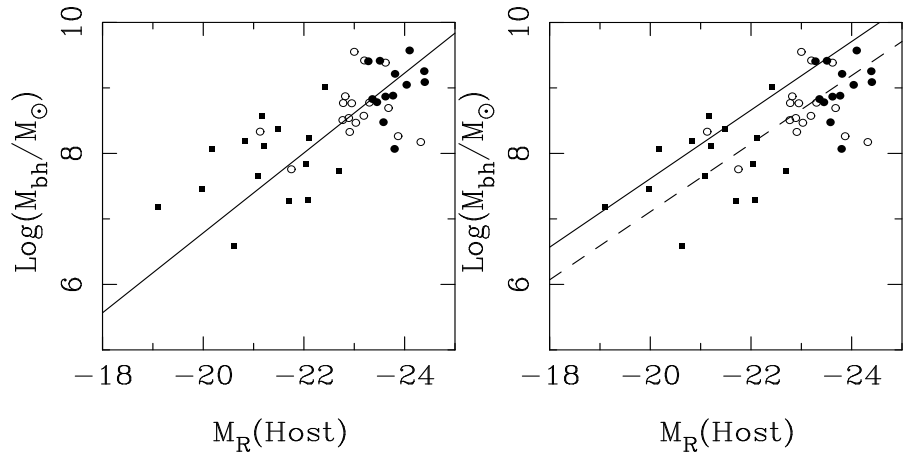,width=12.0cm,angle=0,clip=}}
\caption{Both panels show black-hole mass vs. host galaxy $R$-band
magnitude. The quasars which have black-hole masses estimated from the
$\sbsc{H}{\beta}$ {\sc fwhm} and the $\sbsc{R}{BLR}-\sbsc{L}{5100}$
relation of Kaspi et al. (2000) are shown as open (radio quiet) and filled
(radio loud) circles. The Seyfert galaxies from the sample of Wandel et
al. (1999) which have black-hole estimates derived from reverberation
mapping are shown as filled squares. In the left-hand panel the solid
line is the best fit to the data. The solid line in the right-hand
panel is the predicted relation from Magorrian et al. (1998) 
with $\Mblack/\Mbulge=0.006$. The dashed line in the right-hand panel 
is best fit to the data forcing a constant $\Mblack/\Mbulge$ ratio,
and corresponds to $\Mblack/\Mbulge=0.0025$ (see text for discussion).}  
\label{mainfig}
\end{figure*}

A substantial effort has been invested in providing accurate estimates
of the bulge luminosities for all the objects included in this
study. All 30 objects in the quasar sample have bulge luminosities
derived from two-dimensional modelling of {\sc hst} data from the deep
host-galaxy imaging studies of Dunlop et al. (2000) and Bahcall et
al. (1997). The addition of the Dunlop et al. (2000) host-galaxy
sample means that the number of quasars with reliable bulge
luminosities is a factor of two greater than was available to Laor (1998). 

Of equal importance to this study are revised estimates 
of the
 bulge luminosities within the Seyfert galaxy sample. The bulge
 luminosities used by Wandel (1999) were taken from Whittle
 (1992) and are based upon an application of the Simien \&\ de
 Vaucouleurs (1986) formula relating bulge/disc ratio to galaxy
 morphology. As well as the large amount of scatter associated with
 this empirical formula, it is not clear that it is directly
 applicable to Seyfert galaxies. Indeed, if a typical early-type
 colour of $B-R_{c}=1.6$ is assumed, then the bulge magnitudes
 adopted by Wandel correspond to an average of
 $\sbsc{M}{R}=-22.3\pm0.2$, half a magnitude brighter than the
 recent determination of $\sbsc{M^{\star}}{R}$ by Lin et
 al. (1996). In light of this it seems at least plausible that
 overestimation of the Seyfert galaxy bulge luminosities may
 contribute to the lower $\Mblack/\Mbulge$ ratios found by
 Wandel (1999).

In order to properly quantify the extent to which such overestimates of 
the Seyfert bulge luminosities can explain the Wandel (1999) result,
it is important to obtain bulge luminosity estimates of 
comparable quality to those of the quasar sample. 
Suitable quality {\sc hst} archive data is
available for 9/15 of the Seyfert galaxies, predominantly from the
large-scale Seyfert galaxy imaging study of Malkan, Gorjian \& Tam
(1998). The images of these objects were decomposed into disc and
bulge components in an identical fashion to that utilised in the quasar
host study of Dunlop et al. (2000), and details of the modelling
technique used can be found in McLure, Dunlop \&\ Kukula (2000). The
remaining six Seyfert galaxies have bulge luminosity
estimates taken from the disc/bulge decompositions
of Baggett, Baggett \&\ Anderson (1998) and Kotilainen, Ward
\& Williger (1993). Although determined from ground-based data, these
published bulge luminosities are still expected to be more accurate than
a morphology based estimate. The best estimates of the bulge luminosities 
for the quasar and Seyfert samples are listed in Tables \ref{big} \&
\ref{small} respectively.

\begin{table*}
\begin{center}
\begin{tabular}{cccccccc}
\hline
 Source & $z$ & M$_{\rm R}$(bulge) & $\sbsc{H}{\beta}$ {\sc fwhm}
&Log($\sbsc{L}{ion}$/W)& Log($\Mblack/\Msolar$)&L$_{\rm ion}$/L$_{\rm edd}$\\
 \hline
{\bf RLQ}  &       &                \\
 0137$+$012 & 0.258 & $-$24.0$\pm0.20$ &7.61&38.53&9.05&0.03\\
 0736$+$017 & 0.191 & $-$23.6$\pm0.20$ &2.96&38.89&8.47&0.20\\
 1004$+$130 & 0.240 & $-$24.1$\pm0.75$ &6.34&39.51&9.57&0.07\\
 1020$-$103 & 0.197 & $-$23.4$\pm0.20$ &7.95&38.17&8.83&0.02\\
 1217$+$023 & 0.240 & $-$23.8$\pm0.20$ &3.83&39.15&8.88&0.14\\
 1226$+$023 & 0.158 & $-$24.4$\pm0.75$ &3.52&39.55&9.09&0.22\\
 1302$-$102 & 0.286 & $-$23.5$\pm0.75$ &3.40&39.15&8.78&0.18\\
 1545$+$210 & 0.266 & $-$23.3$\pm0.75$ &7.03&39.14&9.41&0.04\\
 2135$-$147 & 0.200 & $-$23.5$\pm0.20$ &5.50&39.46&9.42&0.09\\
 2141$+$175 & 0.213 & $-$23.8$\pm0.75$ &4.45&39.44&9.22&0.13\\
 2247$+$140 & 0.237 & $-$23.8$\pm0.20$ &2.22&38.66&8.07&0.30\\
 2349$-$014 & 0.173 & $-$24.4$\pm0.75$ &5.50&39.23&9.26&0.07\\
 2355$-$082 & 0.210 & $-$23.6$\pm0.20$ &7.51&38.29&8.87&0.02\\
\hline
{\bf RQQ}  &       &               \\
 0052$+$251 & 0.154 & $-$23.0$\pm0.20$&4.37&38.81&8.77&0.09\\ 
 0054$+$144 & 0.171 & $-$23.6$\pm0.20$&9.66&38.71&9.39&0.02\\
 0157$+$001 & 0.164 & $-$24.3$\pm0.75$&2.14&38.86&8.18&0.37\\
 0204$+$292 & 0.109 & $-$23.3$\pm0.20$&1.04&38.29&8.78&0.03\\
 0205$+$024 & 0.155 & $-$21.1$\pm0.75$&2.90&38.71&8.34&0.18\\
 0244$+$194 & 0.176 & $-$22.8$\pm0.20$&3.70&38.66&8.51&0.11\\
 0923$+$201 & 0.190 & $-$23.3$\pm0.20$&7.31&39.11&9.42&0.04\\
 0953$+$414 & 0.239 & $-$22.8$\pm0.20$&2.96&39.45&8.87&0.29\\
 1012$+$008 & 0.185 & $-$23.9$\pm0.75$&2.64&38.73&7.27&0.22\\
 1029$-$140 & 0.086 & $-$23.0$\pm0.75$&7.50&39.27&9.55&0.04\\
 1116$+$215 & 0.177 & $-$23.7$\pm0.75$&2.92&39.21&8.70&0.26\\
 1202$+$281 & 0.165 & $-$22.8$\pm0.75$&5.01&38.65&8.77&0.06\\
 1307$+$085 & 0.155 & $-$22.9$\pm0.75$&2.36&38.96&8.33&0.33\\
 1309$+$355 & 0.184 & $-$23.0$\pm0.75$&2.94&38.89&8.47&0.20\\
 1402$+$261 & 0.164 & $-$21.8$\pm0.75$&1.91&38.41&7.76&0.34\\
 1444$+$407 & 0.267 & $-$22.9$\pm0.75$&2.48&39.20&8.54&0.35\\
 1635$+$119 & 0.146 & $-$23.2$\pm0.20$&5.10&38.35&8.58&0.05\\
 \hline
\end{tabular}
\end{center}
\caption{Data for the quasar sample. Column three shows the host
galaxy absolute $R$-band bulge magnitudes complete with estimated
errors. Objects which are allocated an error of $\pm0.2$ magnitudes
are all from the Dunlop et al. (2000) study. Objects which are
allocated an error of $\pm0.75$ magnitudes are either highly
morphologically disturbed objects from the Dunlop et al. study, or are
taken from the Bahcall et al. (1997) study. Column four shows the {\sc fwhm} of
the $\sbsc{H}{\beta}$ emission line in units of 1000 kms$^{-1}$. Column
five lists the quasar ionizing luminosity, using the estimate that L$_{\rm
ion}\sim10\lambda$L$_{5100}$. Column six lists the black-hole
masses in solar units as derived from the $\sbsc{H}{\beta}$ {\sc
fwhm} and 5100 \AA \,continuum luminosity (see text). Column seven lists
the quasar ionizing luminosity as a fraction of the Eddington
luminosity.}
\label{big}
\end{table*}

\begin{table*}
\begin{center}
\begin{tabular}{lcccl}
\hline
Source & z & $\sbsc{M}{R}$(bulge) & Log($\Mblack/\Msolar$) &Bulge luminosity details\\
\hline
IC 4329A   &0.016&$-19.11\pm1.00$ &7.18  &B/D decomposition from Kotilainen, Ward \&\ Williger (1993)\\
Fairall 9  &0.047&$-21.49\pm1.00$ &8.38 &B/D decomposition from Kotilainen, Ward \&\ Williger (1993)\\
3C  120    &0.033&$-22.04\pm0.75$ &7.84  &This work, {\sc hst} archive
image, PID=6285, Filter=F675W \\
3C  390.3  &0.056&$-22.42\pm0.75$ &9.01 &This work, {\sc hst} archive
image, PID=5476, Filter=F702W\\
Mrk 79     &0.022&$-20.83\pm0.75$ &8.20 &This work, {\sc hst} archive image, PID=5479, Filter=F606W\\
Mrk 335	   &0.026&$-21.70\pm0.75$ &7.18 &This work, {\sc hst} archive image, PID=5479, Filter=F606W\\
Mrk 509    &0.034&$-22.11\pm1.00$ &8.24 &B/D decomposition from Kotilainen, Ward \&\ Williger (1993)\\
Mrk 590    &0.026&$-22.69\pm0.75$ &7.73 &This work, {\sc hst} archive image, PID=5479, Filter=F606W\\
Mrk 817    &0.032&$-21.20\pm0.75$ &8.12 &This work, {\sc hst} archive image, PID=5479, Filter=F606W\\
NGC 3227   &0.004&$-19.38\pm1.00$ &8.07 &B/D decomposition from Baggett, Baggett \&\ Anderson (1998)\\
NGC 3783   &0.010&$-19.85\pm0.75$ &7.45 &This work, {\sc hst} archive image, PID=5479, Filter=F606W\\
NGC 4051   &0.002&$-20.62\pm1.00$ &6.59 &B/D decomposition from Baggett, Baggett \&\ Anderson (1998)\\
NGC 4151   &0.003&$-19.05\pm1.00$ &7.66 &B/D decomposition from Baggett, Baggett \&\ Anderson (1998)\\
NGC 5548   &0.017&$-21.17\pm0.75$ &8.57 &This work, {\sc hst} archive
image, PID=5479, Filter=F606W\\
NGC 7469   &0.016&$-22.08\pm0.75$ &7.29 &This work, {\sc hst} archive image, PID=5479, Filter=F606W\\
\hline
\end{tabular}
\end{center}
\caption{Data for the Seyfert galaxy sample. Column three shows the
absolute $R$-band bulge magnitudes complete with estimated errors. Objects
whose host galaxies have been modelled for this paper have been
allocated an error of $\pm0.75$ magnitudes. Objects whose bulge
luminosity has been taken from the literature have been allocated an
error of $\pm1.0$ magnitudes. Column four shows the virial black-hole
mass estimates based on reverberation mapping and mean
$\sbsc{H}{\beta}$ {\sc fwhm} measurements from Kaspi
et al. (2000) and the disc model of the BLR discussed in the
text. Column 5 gives details of how the bulge luminosities
were determined. For those objects for which we present new bulge/disc
decompositions from {\sc hst} archive data we have also listed the ID number of
the {\sc hst} proposal from which the data were taken.}
\label{small}
\end{table*}

\section{The black-hole mass vs. bulge luminosity relation}
\label{main}

\begin{figure}
\centerline{\epsfig{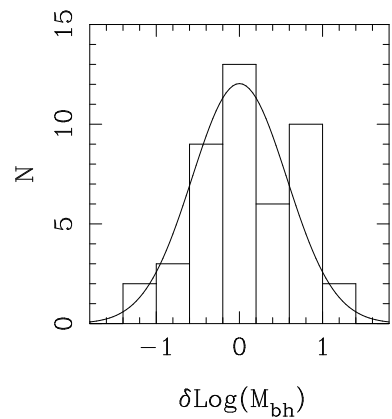}}
\caption{Histogram of the scatter around the best-fitting relation between
black-hole mass and host galaxy magnitude. A gaussian with a standard
deviation of $\sigma=0.59$ is plotted for comparison.}
\label{scatter}
\end{figure}

The black-hole mass vs. bulge luminosity relation for the combined
quasar and Seyfert galaxy sample is plotted in Fig \ref{mainfig}. The
two quantities can be seen to be well correlated, with the
rank-order coefficient of $\sbsc{r}{s}=-0.66$ having a significance of
$4.4\sigma$. In order to obtain a reliable fit to the relation, a 
$\chi^{2}$ minimization was used which properly accounts for the
uncertainties in both the estimated black-hole masses and bulge
luminosities (Press et al. 1992). The best-fitting relation 
($\chi^{2}=64.8$ for 45 d.o.f.) is found to be:
\begin{equation}
\log(\Mblack/\Msolar)=-0.61(\pm0.08){\rm M}_{\rm R}-5.41(\pm1.75)
\end{equation}
\noindent
and is shown as the solid line in the left-hand panel of
Fig \ref{mainfig}.  In terms of black-hole mass the scatter around 
the best-fitting relation of $\sigma=0.59$ is large (see Fig
\ref{scatter}), although still comparable to that found by both 
Magorrian et al. (1998) and Laor
(1998). In contrast to the findings of Wandel (1999) there
is no evidence from Fig \ref{mainfig} that the Seyfert galaxy
$\Mblack-\sbsc{L}{bulge}$ relation is different to that of
quasars. The removal of this discrepancy is due to the 
improved estimates of the Seyfert
galaxy bulge luminosities use here.
As mentioned in Section \ref{bulge}, it was
suspected {\it a priori} that the bulge luminosities adopted 
by Wandel (1999) were probably overestimated. This suspicion is supported by
the mean bulge luminosity derived for the Seyfert galaxies here, 
M$_{\rm R}=-21.0\pm0.3$, which is 1.3 magnitudes fainter than estimated by
Wandel (1999). It appears therefore that the evidence from this
data-set supports a straightforward unification of Seyfert galaxies
and quasars in which black-hole mass scales with bulge
luminosity. In this scenario Fig \ref{mainfig} suggests that a
black-hole mass of $\sim10^{8.5}\Msolar$ needed to power a luminous
quasar requires a host galaxy with a  bulge luminosity of M$_{\rm
R}\sim-23$. This restriction naturally predicts that the hosts of
luminous quasars will be predominantly massive early-type galaxies, in
good agreement with the findings of recent studies (eg. McLure et
al. 1999, Dunlop et al. 2000, Boyce et al. 1998).

\subsection{Do black-hole and bulge mass scale linearly?}
\label{linear}
By combining bulge luminosity with a suitable
mass-to-light ratio it is possible to test whether or not the observed
$\Mblack-\sbsc{L}{host}$ relation is consistent with a
simple linear scaling between black-hole and bulge mass. 
The mass-to-light ratio adopted here is the
${\rm M/L}\propto {\rm L}^{0.31}$ relation determined by 
J$\o$rgensen, Franx \&\ Kj$\ae$rgaard (1996), hereafter JFK96, from their
Gunn-$r$ fundamental plane study. Under the assumption that 
$\Mblack= k\Mbulge$, the $\Mblack-\sbsc{L}{host}$ relation is expected 
to have the form:
\begin{equation}
\log(\Mblack/\Msolar) = \log k -0.52\sbsc{M}{R} -0.66
\label{predict}
\end{equation}
\noindent
where the constant $-0.66$ results from assuming an average bulge colour of 
$M_{r}-M_{R_{c}}=0.37$ (Fukugita et al. 1995) and an absolute Gunn-$r$ 
magnitude for the Sun of $\simeq4.65$ (J$\o$rgensen 1994). A relation of
this form, with a value of $k=0.006$ as determined by Magorrian et
al. (1998), is shown as the solid line in the right-hand panel of
 Fig \ref{mainfig}. It can be seen that, for a given bulge luminosity 
adopting the Magorrian et al. normalization predicts black-hole 
masses approximately 2.5 times larger than determined 
via the $\sbsc{H}{\beta}$ {\sc fwhm} . It is also noteworthy however that
the slope of $-0.52$ expected from a linear $\Mblack-\Mbulge$ relation
is not inconsistent with the best-fitting value of $-0.61\pm0.08$
determined above. Consequently, a least-squares fit of
the data was undertaken with an enforced slope of $-0.52$, and is shown
as the dashed line in the right-hand panel of Fig \ref{mainfig}. This 
can be seen to be a reasonable representation of the data, 
suggesting that the scaling between black-hole and bulge mass is
consistent with being linear. The least-squares fit with an
enforced slope of $-0.52$ has a intercept of $-3.29$ which, when
compared with Equ \ref{predict}, suggests that the constant of
proportionality between black-hole and bulge mass is
$k\simeq0.0025$.

It is worth emphasising that this value for the constant of
proportionality results, at least in part, from our use of a simple,
but realistic disc-like model for the BLR, incorporating the known
orientation biases of optically luminous AGN. If instead one simply uses 
the random orbit relation 
$V = \frac{\sqrt{3}}{2}${\sc FWHM}, then as discussed in Section 2, the
inferred black-hole masses would be $\simeq 3$ times smaller, yielding
the relation $\Mblack = 0.0008 \Mbulge$. 
This is the origin of the findings of Ho
(1999) who studied the $\Mblack-\sbsc{L}{host}$ relation in a sample 16
Seyfert galaxies (12 of which are common to the sample studied here)
using published $B$-band photometry and virial black-hole mass estimates
from reverberation mapping and $\sbsc{H}{\beta}$ {\sc fwhm}
measurements. Ho found that the reverberation
mapping black-hole estimates were lower than expected from the
$\Mblack-\sbsc{L}{host}$ relation, suggesting a value of 
$k\simeq0.001$. 
Further evidence that simple random-orbit calculations for the BLR  
lead to systematic underestimation of black-hole mass is 
discussed in Section 5.

\subsection{The ionizing continuum luminosity vs. bulge luminosity correlation}

\begin{figure}
\centerline{\epsfig{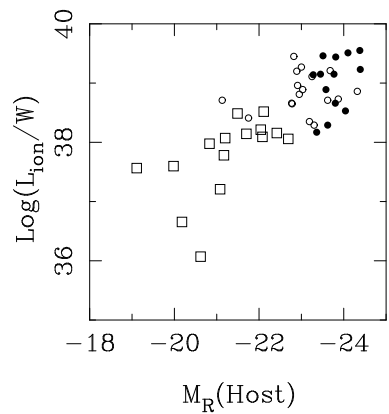}}
\caption{Ionizing continuum luminosity vs. host-galaxy $R$-band
magnitude. The symbols are the same as Fig \ref{mainfig}.}
\label{lion}
\end{figure}
Fig \ref{lion} shows the correlation between bulge luminosity and
continuum ionizing luminosity for the full sample, where we have
adopted the relation $\sbsc{L}{ion}\simeq10\lambda\sbsc{L}{5100}$
which was found to be the average of a large scatter by Wandel, 
Peterson \&\ Malkan (1999). These two quantities are
well correlated ($\sbsc{r}{s}=-0.73, 4.85\sigma$), in apparent
agreement with the results of the combined Seyfert galaxy and quasar study
of McLeod \&\ Rieke (1995). McLeod \&\ Rieke
found a relation between $H$-band host luminosity and $B$-band nuclear
luminosity in the form of a lower limit, such that 
$\sbsc{M}{H}$(host)$\ge\sbsc{M}{B}$(nucleus).

However, it is also noticeable from Fig \ref{lion} that the correlation
between $\sbsc{L}{ion}$ and $\sbsc{L}{host}$ is entirely dependent
upon the low luminosity Seyfert galaxies, with the quasar sample on 
its own displaying no significant correlation, $\sbsc{r}{s}=0.27,
1.45\sigma$. The conclusion that the correlation between
$\sbsc{L}{ion}$ and $\sbsc{L}{host}$ breaks down at higher nuclear
luminosities is supported by the results of a study of high-luminosity 
quasars by Percival et al. (2000). Although including
quasars up to a factor of ten more luminous than those studied here,
Percival et al. find very similar host galaxy luminosities. The
suggestion is therefore that high luminosity quasars operate at higher
accretion rates, rather than harbouring significantly more massive 
black-holes. 

The correlation between $\sbsc{L}{ion}$ and $\sbsc{L}{host}$ amongst
the Seyfert galaxies raises the possibility that the observed
correlation between $\Mblack$ and $\sbsc{L}{host}$  could in fact be
simply due to independent correlations with $\sbsc{L}{ion}$. In order
to investigate this possibility the partial
Spearman rank correlation test (Macklin 1982) has been used to
quantify the correlation between $\Mblack$ and
$\sbsc{L}{host}$, independent of $\sbsc{L}{ion}$. The correlation 
between $\Mblack$ and $\sbsc{L}{host}$ at constant $\sbsc{L}{ion}$ has 
a coefficient of $\sbsc{r}{s}=-0.28$ which has a significance of
$1.9\sigma$. Therefore,
although the $\Mblack$ and $\sbsc{L}{host}$ correlation is still clearly
present, its significance is substantially lowered. 

Unfortunately, due to the heterogeneous nature of the sample of Seyfert
 galaxies which currently have reverberation mapping data available,
 it is difficult to quantify properly the significance of the 
$\Mblack-\sbsc{L}{host}$ relation. In order to overcome this problem, a
 substantially larger sample of AGN is required, selected from a
 narrow slice in both nuclear luminosity and redshift. 
\subsection{Black-hole mass and unification}

\begin{figure}
\centerline{\epsfig{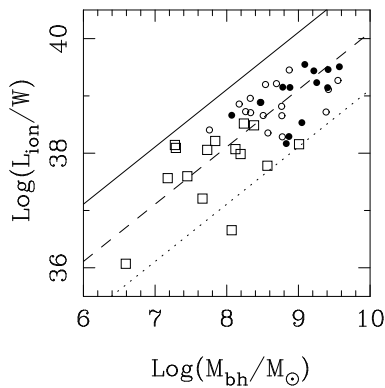}}
\caption{Ionizing continuum luminosity vs. black-hole mass. The three
lines indicate the location of objects radiating with a luminosity equal to
$\sbsc{L}{Edd}$ (solid), $0.1\sbsc{L}{Edd}$ (dashed) and
$0.01\sbsc{L}{Edd}$ (dotted). The symbols are the same as Fig \ref{mainfig}.}
\label{ledd}
\end{figure}
The black-hole mass estimates for the 17 radio-quiet and 13 radio-loud 
quasars studied here provide an opportunity to determine the
influence (if any) of black-hole mass on quasar radio luminosity. The
ionizing continuum luminosity distributions of the two quasar
sub-samples are indistinguishable, with a Kolmogorov-Smirnov (KS) test 
returning a probability of $p=0.18$ that both sub-samples are drawn from the 
same underlying distribution. Consequently, any difference in the
distribution of black-hole masses between the two sub-samples is
presumably linked to the difference in radio properties. 

With the results of
Section \ref{linear} showing that AGN black-hole
and bulge masses are consistent with being directly proportional, 
it is expected that the $0.77$ magnitude difference in median 
bulge luminosity between the quasar sub-samples should 
translate directly into a corresponding 
difference in average black-hole mass. This is indeed
the case, with the median black-hole mass of the
radio-loud sub-sample a factor of three larger than that of the
radio-quiet sub-sample. A  natural
division between the quasar sub-samples appears to occur at
$\Mblack\sim10^{8.8}\Msolar$. Only 2/13 of the radio-loud quasars have
$\Mblack<10^{8.8}\Msolar$, while only 4/17 of the radio-quiet have
$\Mblack>10^{8.8}\Msolar$. This difference in black-hole mass 
distributions is shown to be significant at the $2.9\sigma$ ($p=0.004$) 
level by a KS test. The implication from the
quasar sample is therefore that (albeit with substantial overlap) 
for a given nuclear luminosity, the
probability of a source being radio-loud increases with black-hole
mass, or alternately decreases with ${\rm L_{ion}}/\sbsc{L}{Edd}$. 

Fig \ref{ledd} shows a plot of estimated ionizing luminosity
vs. black-hole mass, together with lines indicating the location of 
objects radiating at 100\%, 10\% and
1\% of the Eddington limit. Two aspects of this figure are
noteworthy. Firstly, it can be seen that the vast majority of the
objects appear to be radiating in the range $5\rightarrow30\%$ of
the Eddington limit, with a mean value of ${\rm L_{ion}}/{\rm
L_{edd}}=0.15$. This is despite the fact that the sample spans
a range in ionizing luminosity of four orders of magnitude, and is in
contrast to the results of Wandel, Peterson \& Malkan (1999) and Kaspi
et al. (2000) who both find that the ${\rm L_{ion}}/{\rm
L_{edd}}$ ratio increases weakly with continuum
luminosity. Secondly, there appears to be
no evidence that the Seyfert galaxies are operating in a different
accretion regime to the quasars, contrary to the finding of 
Lu \& Yu (1999). However, it should also perhaps be noted that
differences between the UV spectra of Seyfert galaxies and quasars could
well mean that a global correction of the form
$\sbsc{L}{ion}=10\lambda\sbsc{L}{5100}$ is inappropriate.

\section{The normalization of the ${\bf \Mblack-\Mbulge}$ relation}
\label{fp}

 In this section the possibility of calibrating the virial
$\sbsc{H}{\beta}$ {\sc fwhm} black-hole estimates, by taking advantage
of the remarkably small scatter in the recently discovered
$\Mblack-\sigma$ correlation, is explored. Considering that $\sbsc{H}{\beta}$ 
{\sc fwhm} measurements can be obtained with relative ease for
powerful AGN, in comparison with determining the host-galaxy stellar 
velocity dispersion, this is a question of some interest. 

Although stellar velocity measurements are not available for the
quasar sample, a comparison is still possible for the 19 objects
which are taken from the host-galaxy study of McLure et al. (1999) and
Dunlop et al. (2000). One of the main successes of this study was the
first demonstration that the massive elliptical host galaxies of these 
quasars follow an identical Kormendy relation ($\mu_{e}-r_{e}$) to
normal inactive ellipticals. Therefore, because the Kormendy relation
is simply a projection of the fundamental plane that elliptical
galaxies occupy in the three dimensional space defined by
$(r_{e},\sigma, I_{e})$, it is possible to use the $r_{e}$ \&\ $I_{e}$
data for these objects to estimate their stellar
velocity dispersions. The fundamental plane parameters adopted here
are those determined for the Coma cluster by JFK96:
\begin{equation}
\log r_{e}=1.31\log \sigma -0.84 \log<I>_{e}-0.0082
\end{equation}
\noindent
Although there is evidence that the form of the fundamental plane changes
with redshift, the form of this evolution
($\Delta \log({\rm M/L})=-0.26\Delta z$; J$\o$rgensen et al. 1999)
should not introduce significant errors at the average redshift of the 
quasar sample ($z\simeq0.2$). 

Due to the disagreement over what the correct form of the
$\Mblack-\sigma$ relation is, both the $\Mblack \propto \sigma^{3.75}$ 
(Gebhardt et al. 2000a) and $\Mblack \propto \sigma^{4.72}$ (Merritt
\&\ Ferrarese 2000) versions of the relation
 have been used to convert the velocity dispersion estimates 
into black-hole masses. In both cases
the $\Mblack-\sigma$ relations predict black-hole masses which are
in remarkably good agreement with our virial $\sbsc{H}{\beta}$ estimates, with
median $\Mblack(\sigma)/\Mblack(\sbsc{H}{\beta})$ ratios of 1.04 \&\
1.31 for the Gebhardt et al. and Merritt \&\ Ferrarese relations
respectively. The clear implication of this comparison is that the use of
a disc-like BLR, combined with limits of disc orientation relative to the
observer, has led to reliable virial $\sbsc{H}{\beta}$ {\sc fwhm}
estimates 
of black-hole mass. In contrast, the standard assumption of purely
random BLR velocities lead to virial $\sbsc{H}{\beta}$ {\sc fwhm} 
black-hole mass estimates which appear to be systematically low by a 
factor of $2\rightarrow3$. 

A similar systematic problem with the assumption of random orbits 
was alluded to by Gebhardt
et al. (2000b) when they compared (random orbit) 
virial black-hole mass estimates, based on
reverberation mapping and $\sbsc{H}{\beta}$ {\sc fwhm} measurements, 
with velocity dispersion black-hole masses for a sample of seven Seyfert
galaxies (five of which are common to the sample studied here). Although
Gebhardt et al. (2000b) concluded that the discrepancy between the two
black-hole mass estimates was not significant, they did comment that
an increase in the reverberation mapping estimates by a factor of
$\simeq2$ would improve the agreement. 

Finally, in further support of the relation
$\Mblack \simeq 0.0025\Mbulge$ derived from our $\sbsc{H}{\beta}$ {\sc fwhm}
analysis, we note that Gebhardt et al. (2000b) also 
comment that new three integral models show that the black-hole masses 
originally determined by Magorrian et al. (1998) were overestimated by 
a factor of $\sim3$, again leading to the conclusion that 
$\Mblack\sim 0.002 \Mbulge$. 

\section{Conclusions}

The black-hole masses for a sample of 30 optically-matched radio-loud
and radio-quiet quasars have been estimated using $\sbsc{H}{\beta}$
{\sc fwhm} and continuum luminosity measurements from new and
previously published nuclear spectra. Reliable quasar bulge
luminosities have been combined with new and published disc/bulge
decompositions for a sample of Seyfert galaxies with reverberation
mapping black-hole estimates, in order to study the form of the
$\Mblack-\sbsc{L}{host}$ relation over a large baseline in AGN
luminosity. The results of the $\sbsc{H}{\beta}$ black-holes mass
estimates have been compared with the $\Mblack-\sigma$ correlation in an
attempt to check the normalization of the $\Mblack-\Mbulge$
relation. The main conclusions of this study can be summarized as
follows:
\begin{enumerate}
\item{Host-galaxy bulge luminosity and black-hole mass are found to be
well correlated, albeit with scatter of 0.6 dex in $\Mblack$. Assuming a 
simple but realistic disc-like geometry for the BLR, coupled with limits on
viewing angle for optically-luminous AGN, the form
of the $\sbsc{L}{host}-\Mblack$ correlation is shown to be consistent
with $\Mblack\simeq0.0025\Mbulge$.}
\item{Contrary to the results of Wandel (1999) no evidence is found 
that Seyfert galaxies follow a different $\Mblack-\sbsc{L}{host}$ 
relation to quasars. The results presented here suggest that
unification of Seyfert galaxies and radio-quiet quasars via a simple
scaling in black-hole mass is tenable.} 
\item{A comparison of the virial black-hole estimates based on
$\sbsc{H}{\beta}$ {\sc fwhm} measurements with
those predicted by the $\Mblack-\sigma$
relation suggests that our $\sbsc{H}{\beta}$ black-hole estimates are, on
average, accurate to within a factor $< 1.5$. 
These two, completely independent approaches to black-hole mass estimation
both support a black-hole mass vs. bulge mass relation
of the form $\Mblack\simeq0.0025\Mbulge$.}
\item{The distribution of black-hole masses in the radio-quiet and
radio-loud quasar sub-samples are found to be significantly different
($p=0.004$), with the median radio-loud black-hole mass a factor of
three larger than the equivalent radio-quiet value. It appears that, for
some reason, a radio-loud quasar requires a black-hole mass 
$> 6\times10^{8}\Msolar$.} 
\end{enumerate}
\noindent
Finally, perhaps the most important conclusion resulting from this study is
that easily obtainable AGN nuclear spectra can be be used to provide
black-hole mass estimates which are accurate to within a factor of
$< 2$. 
It is therefore possible that large-scale studies of AGN black-hole
demography can be undertaken without the need for more demanding and 
time-intensive stellar velocity dispersion measurements.
\begin{figure*}
\centerline{\epsfig{file=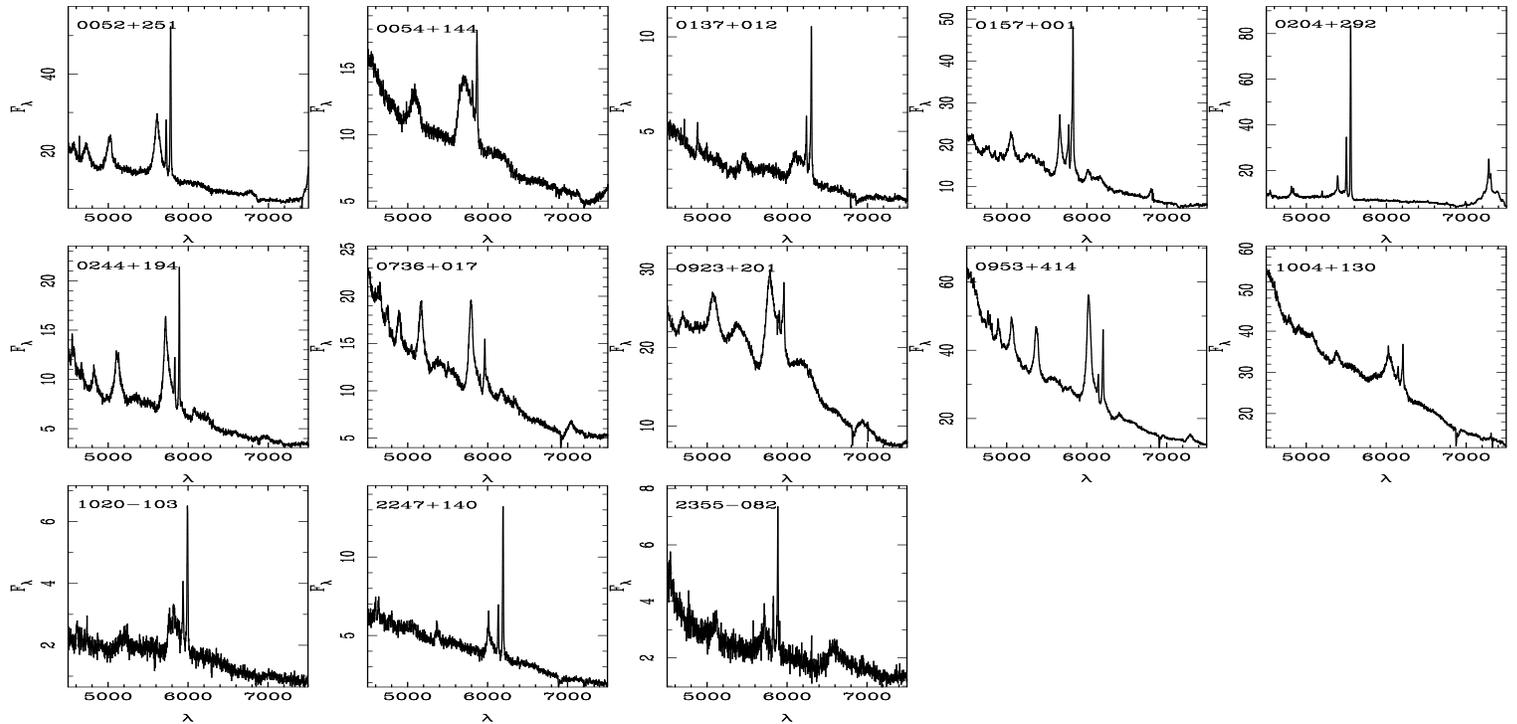,width=20.0cm,angle=0,clip=}}
\caption{The final reduced quasar spectra obtained with IDS 
on the Isaac Newton Telescope. The spectra are displayed in the
observed frame with wavelength in Angstroms. The vertical axis shows
object flux in units of $10^{-16}$ergs\,cm$^{-2}$s$^{-1}$\AA$^{-1}$. 
Object names are displayed in the top left-hand corner of each
individual spectrum.}
\label{spectra}
\end{figure*}
\section{acknowledgements}
The INT is operated on the island of La Palma by the Isaac Newton
Group in the Spanish Observatorio del Roque de los Muchachos of the
Instituto de Astrofisica de Canarias. Based on observations with
the NASA/ESA Hubble Space Telescope, obtained
at the Space Telescope Science Institute, which is operated by the
Association of Universities for Research in Astronomy, Inc. under NASA
contract No. NAS5-26555.
This research has made use of the NASA/IPAC Extragalactic Database (NED)
which is operated by the Jet Propulsion Laboratory, California Institute
of Technology, under contract with the National Aeronautics and Space
Administration. RJM acknowledges a PPARC PDF.
\section{references}
Antonucci R., Miller J.S., 1985, ApJ, 297, 621\\
Baggett W.E., Baggett S.M., Anderson K.S.J., 1998, AJ, 116, 1626\\
Bahcall J.N., Kirhakos S., Saxe D.H., Schneider D.P., 1997, ApJ, 479, 642\\
Boroson T.A., Green R.F., 1992, ApJS, 80,109\\
Boyce P.J., et al., 1998, MNRAS, 298, 121\\
Corbin M.R., 1997, ApJS, 113, 245\\
Dunlop J.S., et al., 2001, MNRAS, submitted, \\
Fukugita M., Shimasaku K., Ichikawa T., 1995, PASP, 107, 945\\
Gebhardt K., et al., 2000a, ApJ, 539, L13\\
Gebhardt K., et al., 2000b, ApJ, 543, L5\\
Ho, L.C., 1999, in Chakrabarti S.K., ed, Proc Observation evidence for
the black holes in the universe, Kluwer, Dordrecht, p.157\\
J$\o$rgensen I., 1994, PASP, 106, 967\\
J$\o$rgensen I., Franx M., Kj$\ae$rgaard P., 1996, MNRAS, 280, 167\\
J$\o$rgensen I., Franx M., Hjorth J., van Dokkum P.G., 1999, MNRAS, 308, 833\\
Kaspi S., Smith P.S., Netzer H., Maoz D., Jannuzi B.T., Giveon U.,
2000, ApJ, 533, 631\\
Kauffmann G., Haehnelt M., 2000, MNRAS, 311, 576\\
Kormendy J., Richstone D., 1995, ARA\&A, 33, 581\\
Kotilainen J.K., Ward M.J., Williger, 1993, MNRAS, 263, 655\\
Laor A., 1998, ApJ, 505, L83\\
Lin H., Kirshner P.P., Schectman S.A., Landy S.D., Oemler A., Tucker
Lu Y., Yu Q., ApJ, 1999, 526, L5\\
D.L., Schechter P.L., 1996, ApJ, 464, 60\\
Macklin J.T., 1982, MNRAS, 199, 1119\\
McLeod K.K., Rieke G.H., 1995, ApJ, 441, 96\\
McLure R.J., Dunlop J.S., Kukula M.J.,Baum S.A., O'Dea C.P., Hughes
D.H., 1999, MNRAS, 308, 377\\
McLure R.J., Dunlop J.S., Kukula M.J., 2000, MNRAS, 318, 693\\
Magorrian J., et al., 1998, AJ, 115, 2285\\
Malkan M., Gorjian V., Tam R., 1998, ApJS, 117, 25\\
Merritt D., Ferrarese L., 2000, MNRAS, 2001, 320, L30\\
Percival W.J., Miller L., McLure R.J., Dunlop J.S., 2000, MNRAS,
in press, astro-ph/0002199\\
Press W.H., Teukolsky S.A., Vetterling W.T., Flannery B.P., 1992,
Numerical Recipes, Cambridge University Press\\
Schade D.J., Boyle B.J., Letawsky M., 2000, MNRAS, 315, 498\\
Silk J., Rees M.J., 1998, A\&A, 331, L1\\
Simien F., de Vaucouleurs G., 1986, ApJ, 302, 564\\
van der Marel, R.P., 1999, in Barnes J.E., Sanders D.B., eds, Proc
IAU Symp. 186, Galaxy interactions at low and high redshift, Kluwer,
Dordrecht, p. 333\\
Vestergaard M., Wilkes B.J., Barthel P.D., 2000, ApJ, L103\\ 
Wandel A., 1999, ApJ, 519, L39\\
Wandel A., Peterson B.M., Malkan M.A., 1999, ApJ, 526,579\\
Wang Y.P., Biermann P.L., Wandel A., 2000, A\&A, in press,
astro-ph/0008105\\
Wills B.J., Browne I.W.A., 1986, ApJ, 302, 56\\
Wilman R.J., Fabian A.C., Nulsen P.E.J., 2000, MNRAS, in press,
astro-ph/0008019\\
Whittle M., 1992, ApJS, 79, 49\\
\end{document}